\documentclass[prb,reprint,amsmath,amssymb,showpacs,superscriptaddress,floatfix,longbibliography]{revtex4-1}
\usepackage{graphicx,epsfig}% Include figure files
\usepackage{amsmath}% bold math
\usepackage{xcolor}% coloured text
\usepackage[breaklinks=true,colorlinks,citecolor=blue,linkcolor=blue,urlcolor=blue]{hyperref}
\usepackage{subfigure}
\usepackage{color}
\usepackage[capitalize]{cleveref}
\def\be{\begin{equation}}
\def\ee{\end{equation}}
\def\nn{\nonumber}
\def\bea{\begin{eqnarray}}
\def\eea{\end{eqnarray}}

\begin{document}

\title{Anisotropic plasmons, excitons and electron energy loss spectroscopy of phosphorene}
\author{Barun Ghosh}
\affiliation{Dept. of Physics, Indian Institute of Technology Kanpur, Kanpur 208016, India }
\author{Piyush Kumar}
\affiliation{Dept. of Electrical Engineering, Indian Institute of Technology Kanpur, Kanpur 208016, India}
\author{Anmol Thakur}
\affiliation{Dept. of Physics, Indian Institute of Technology Kanpur, Kanpur 208016, India }
\author{Yogesh Singh Chauhan}
\affiliation{Dept. of Electrical Engineering, Indian Institute of Technology Kanpur, Kanpur 208016, India}
\author{Somnath Bhowmick}
\affiliation{Dept. of Materials Science and Engineering, Indian Institute of Technology Kanpur, Kanpur 208016, India}
\date{\today}
\author{Amit Agarwal}
\email{amitag@iitk.com}
\affiliation{Dept. of Physics, Indian Institute of Technology Kanpur, Kanpur 208016, India }

\begin{abstract}
In this article, we explore the anisotropic electron energy loss spectrum (EELS) in monolayer phosphorene based on ab-initio time dependent density functional theory calculations. Similar to black phosphorous, the EELS of undoped monolayer phosphorene is characterized by anisotropic excitonic peaks for energies in vicinity of the bandgap, and by interband plasmon peaks for higher energies. On doping, an additional intraband plasmon peak also appears for energies within the bandgap. Similar to other two dimensional systems, the intraband plasmon peak disperses as $\omega_{\rm pl} \propto \sqrt{q}$ in both the zigzag and armchair directions in the long wavelength limit, and deviates for larger wavevectors. The anisotropy of the long wavelength plasmon intraband dispersion is found to be inversely proportional to the square root of the ratio of the effective masses: $\omega_{\rm pl}(q \hat{y})/\omega_{\rm pl}(q \hat{x}) = \sqrt{m_x/m_y}$. 
\end{abstract}

% 
%\pacs{}
%
\maketitle

\section{Introduction}

Within the family of 2D materials\cite{Geim:2007aa,graphene_tr,MoS2_tr,2D_review,Dubertret:2015aa,phosphorene1,BP-transistor,Ghosh:2015,Novoselovaac9439,Nahas:2016,Mardanya:2016}, phosphorene (few layers of black phosphorous)  - a direct bandgap semiconductor with a puckered atomic structure, has a highly anisotropic band-structure \cite{Li:2014,Liu:2014,Ling14042015}. 
It has a massive Dirac like energy dispersion along the armchair and a parabolic Schr\"odinger like energy dispersion along the zigzag direction\cite{Qiao:2014,Rodin:2014,Rudenko:2014,Ezawa:2014}. This results in highly anisotropic single particle electronic\cite{Qiao:2014}, thermal\cite{C5NR01821H}, many particle excitonic\cite{Tran_2014} and plasmonic properties \cite{Low}. In particular, the plasmon dispersion has  been recently calculated based on low energy continuum Hamiltonian, and it is found to be highly anisotropic\cite{Low,Rodin} with different doping dependence in different directions, depending on the number of layers.  

From an experimental viewpoint, momentum resolved electron energy loss spectroscopy\cite{EELS_review} (EELS) directly probes the loss function of a material, which in turn is simply the inverse of the imaginary part of the dynamical interacting dielectric constant of a material: ${\cal E}_{\rm Loss}({\bf q},\omega) = - \Im  [1/\epsilon_M({\bf q},\omega)]$. It has been used extensively in a variety of materials such as graphene\cite{Eberlein,Shin,Kinyanjui:2012,Tegenkamp:2011,Langer:2013,Liou:2015,Kramberger:2008,Yan2011}, transition metal dichalcogenides \cite{Coleman, Priya}, and bulk black phosphorous\cite{Schuster}, to explore the single particle and collective excitations such as excitons and plasmons. In terms of computational methods, calculations based on effective low energy  continuum \cite{Wunsh:2006,Agarwal2:2015} as well as the tight-binding Hamiltonian \cite{Hill,Lam,PRB.92.115440} are very insightful, but they fail to capture many of the experimental aspects of the EELS spectrum accurately; in particular the low energy intraband plasmons at large wave-vectors and the high energy interband plasmons involving transitions across various energy bands.  For example, in doped graphene, the low energy\cite{Wunsh:2006,Hwang:2007, Agarwal:2015} and tight-binding approach\cite{Hill} fails to capture the plasmon anisotropy at finite wave-vectors in the $\Gamma-K$ and the $\Gamma-M$ direction, which has been  observed experimentally \cite{Tegenkamp:2011,Langer:2013}. However, the loss function and corresponding plasmon dispersion relation is generally well described by density functional theory based {\it ab-initio} calculations \cite{Andersen:2013,Vacacela:2016,Johari:2011,Despoja:2013,Kaltenborn:2013,Anderson:2014,Trevisanutto:2008,Yan2011}. In case of  phosphorene, the low energy intraband plasmon spectrum has been studied using the continuum approximation \cite{Low,Rodin}. Further, the effects of strain\cite{Lam} and disorder\cite{PRB.92.115440} on the plasmon spectrum have been incorporated using a tight-binding approach, but to the best of our knowledge, there is no \textit{ab-initio} based study of the anisotropic plasmon dispersion (both low as well as high energy) and the high energy EELS spectrum of monolayer phosphorene.
Motivated by this, we present an \textit{ab-initio} density functional theory (DFT) based study of the anisotropic EELS spectrum, which includes low energy intraband plasmons at finite doping, excitons and interband plasmons in monolayer phosphorene.  

We find that the crystal anisotropy of bulk black phosphorous is preserved down to it's single layer, leading to a highly anisotropic electronic band structure, which results in a direction dependent EELS. In case of finite doping, we find  an intraband plasmon mode which lies well below the band gap of phosphorene. Interestingly, it has a $\sqrt{q}$ dispersion for small wave vector in each of the two principal direction (parallel to the armchair and the zigzag edge of monolayer phosphorene), which signifies the two dimensional nature of the plasmon mode. We also find a highly dispersive mode in the EELS, which appears at slightly higher energy than the bandgap of phosphorene; it is identified as the exciton peak. Interestingly, while the exciton peak appears in the armchair direction,  it is completely absent along the zigzag edge.  The other high energy peaks correspond to different interband transitions, with a very distinct peak appearing for energies close to $5$ eV. We also observe a general trend that with increasing momentum transfer, all the resonant features (excitations) of the EELS spectrum are blue shifted and they gradually loose their strength.

The paper is organized as follows: in Sec.~\ref{Sec2} we discuss the formulation for calculating the interacting density response function and the corresponding EELS spectrum, along with the computational details of our \textit{ab-initio} study to get the electronic band structure of phosphorene. Next, we describe the calculated EELS spectrum in Sec.~\ref{Sec3}, followed by a detailed discussion focused on anisotropy of intraband plasmons in Sec.~\ref{sec:plasmon}. Finally, we summarize our findings in Sec.~\ref{Sec5}.

\section{Theory and Computational details}
\label{Sec2}
\subsection{Dynamical dielectric and loss function }
\label{sec2}
Our starting point is the non-interacting density-density response function $( \chi^0_{\bf{GG'}} )$ for a periodic lattice, which is obtained from the  Adler-Wiser  formula given by\cite{PhysRev.126.413,PhysRev.129.62},
\begin{eqnarray} \label{eq:chi0}
& & \chi^0_{\mathbf{G} \mathbf{G}^{\prime}}(\mathbf{q}, \omega) = \frac{1}{\Omega}
\sum_{\mathbf{k}}^{\mathrm{BZ}} \sum_{n, n^{\prime}}
\frac{f_{n\mathbf{k}}-f_{n^{\prime} \mathbf{k} + \mathbf{q} }}{\omega + \epsilon_{n\mathbf{k}} - \epsilon_{n^{\prime} \mathbf{k} + \mathbf{q} } + i\eta}  \times \\
& & \langle \psi_{n \mathbf{k}} | e^{-i(\mathbf{q} + \mathbf{G}) \cdot \mathbf{r}} | \psi_{n^{\prime} \mathbf{k} + \mathbf{q} } \rangle_{\Omega_{\mathrm{cell}}}
\langle \psi_{n\mathbf{k}} | e^{i(\mathbf{q} + \mathbf{G}^{\prime}) \cdot \mathbf{r}^{\prime}} | \psi_{n^{\prime} \mathbf{k} + \mathbf{q} } \rangle_{\Omega_{\mathrm{cell}}}~. \nonumber
\end{eqnarray} 
The Kohn-Sham energy eigenvalues $\epsilon_{n\bf{k}}$,the wave function $\psi_{n\bf{k}}$ and the corresponding Fermi-Dirac occupation function $f_{n\bf{k}}$ for the n$^{\rm th}$ band at wave vector $\bf{k}$ are obtained from the ground state calculations performed using the density functional theory.

Within the framework of time dependent density functional theory (TDDFT) the interacting density-density response function can be obtained by solving a Dyson-type equation.  
Expanding in a plane wave basis (valid for a periodic system), the interacting response function can be expressed as 
\begin{eqnarray}
\chi_{\bf{GG'}}(\bf{q},\omega) &= &\chi_{\bf GG'}^0(\bf{q},\omega) \\
& +& \sum_{\bf{G_1,G_2}}\chi_{\bf{GG_1}}^0({\bf q},\omega) K_{\bf{G_1G_2}}(q)\chi_{\bf{GG_2}}(\bf{q},\omega), \nonumber
\end{eqnarray}
where $\bf{G}$ and $\bf{q}$ are the reciprocal lattice vector and the wave vector, respectively and $K_{\bf{G_1G_2}}$ is the (2D truncated for monolayer phosphorene) Coulomb kernel.  
The exchange-correlation part of the kernel is neglected within the framework of the random phase approximation (RPA). 
Using the $\chi_{\bf{G}\bf{G'}}$ matrix, 
the macroscopic dielectric function ($\epsilon_M$) is obtained as,
\begin{equation}
\epsilon_M^{-1}({\bf q },\omega) = 1 -v_c(\bf{q})\chi_{\bf{G=0G'=0}}(\bf{q},\omega)
\end{equation}
where $v_c(\bf{q})$ is the (2D truncated for monolayer phosphorene) Coulomb Kernel as described in Ref.~[\onlinecite{PhysRevB.73.205119}]. The dynamical loss function, which is directly related to the EELS, is calculated as 
\begin{equation}
{\cal E}_{\rm Loss}({\bf q},\omega)=-\Im [\epsilon_M^{-1}(q,\omega)]=v_c(\bf{q}) \Im [\chi_{\bf{G}=0,\bf{G'}=0 }(\bf{q},\omega)].
\end{equation}
Plasmons (collective density excitations) are the characterized by the zeroes of the real part of the macroscopic dielectric function (the denominator of the density-density response function within RPA).  
\begin{figure}
\includegraphics[width=1.0 \linewidth]{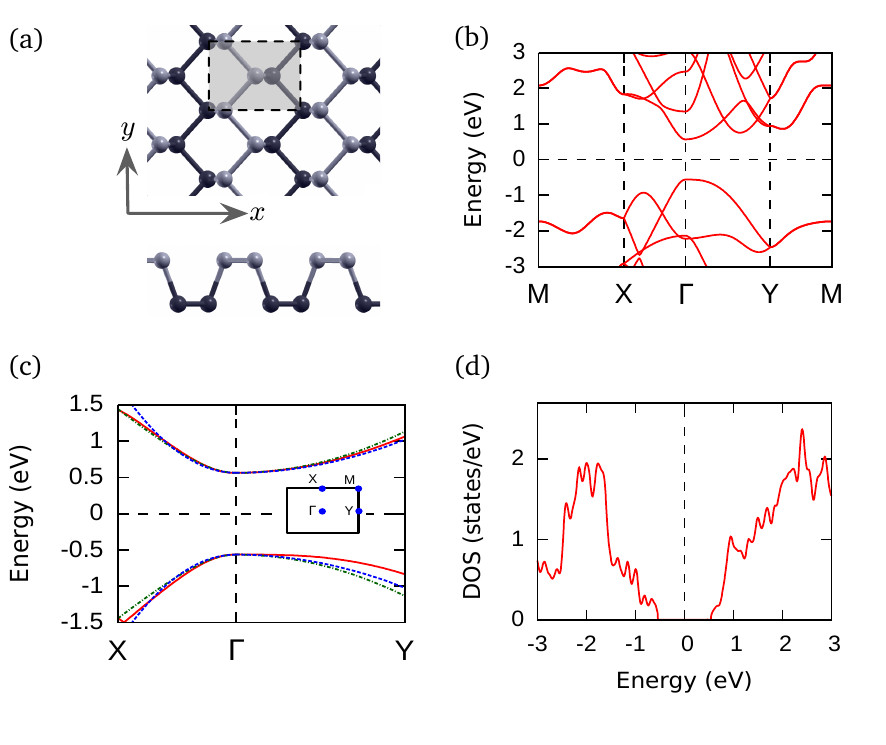}
\caption{(a) Geometric structure of phosphorene monolayer: top and side view with the shaded region indicating the unit cell. (b) The electronic band structure of phosphorene monolayer along the high symmetry axes, calculated using the GLLB-SC functional. (c) The comparison of the effective low energy Hamiltonian in Eq.~\eqref{disp0} (green dashed line) and Eq.~\eqref{disp2} (blue dashed line), with the DFT bandstructure (red line) in vicinity of the $\Gamma$ point. (d) The corresponding density of states.}
\label{bs_dos}
\end{figure}

\subsection{Computational details}
\label{sec3}
Electronic band structure calculations are performed using density functional theory, as implemented in the GPAW package \cite{PhysRevB.71.035109,0953-8984-22-25-253202,ISI:000175131400009}. 
The kinetic energy cut-off for the plane wave basis set is taken to be 500 eV. Initially, all the atomic positions are relaxed (using GGA) until the forces on each atom are less than 0.001 eV/\AA. Next, we calculate the single particle states [to be used as input for evaluating the response function in Eq.~\eqref{eq:chi0}] using the GLLB-SC functional which includes the GLLB type exchange proposed by Gritsenko et al. and PBEsol correlation, which has been found to improve the bandgap in case of semiconductors\cite{PhysRevA.51.1944}. While a k-point grid of $16\times 22\times 6$ ( $16\times 22\times 1$ ) is used for the Brillouin zone integrations of the bulk (monolayer) structure, for the electronic band structure calculation, a much denser k-point grid of 32 $\times$ 44 $\times$ 14 ( 64 $\times$ 88 $\times$ 1) is used for calculating the $q$ dependence of the EELS, giving a momentum resolution of $\sim$0.043 \AA$^{-1}$ and $\sim$0.0215 \AA$^{-1}$ respectively for the bulk and the monolayer phosphorene. In case of the latter, we use a vacuum layer of 20 \AA~ in the direction perpendicular to the phosphorene plane to suppress any interaction between two replica images in the vertical direction.

EELS calculations are performed using the generalized random phase approximation (RPA). The generalized RPA uses the local field factors to add the impact of the exchange and correlation effects to the Hartree field \cite{GV}. Because of it's long-range nature, the Coulomb potential of one layer can interact with its periodic replicas, which is avoided by taking a 2D truncated Coulomb kernel, following Ref~[\onlinecite{PhysRevB.73.205119}]. We consider up to 50 empty bands to correctly describe all the electronic excitations. For the local field corrections, a cut-off energy of 50 eV is used for \textbf{G} and ${\rm \textbf{G}}^{\prime}$ vectors, which corresponds to 259 (135) plane waves for monolayer phosphorene (bulk black phsophorous). Doping or the change in carrier concentration is achieved by shifting the position of the Fermi energy ($E_{\rm f}$).

\subsection{Electronic band structure of phosphorene}
\label{sec4}
Phosphorene has a layered structure with each phosphorus atom covalently bonded with three adjacent atoms, forming a $sp^3$ hybridized puckered honeycomb structure [see Fig \ref{bs_dos}(a)]. As shown in the figure, mutually perpendicular armchair and zigzag directions are aligned along the $x$ and $y$ axis, respectively. The lattice parameters of the bulk unit cell are found to be $a=4.56$ \AA~ (armchair), $b=3.31$ \AA~ (zigzag) and $c=11.30$ \AA,~ which are in good agreement with the values reported in the literature\cite{Liu:2014}. As shown in Fig \ref{bs_dos}(a), $\Gamma-X$ and $\Gamma-Y$ are the high symmetry directions in the reciprocal lattice, which are aligned along the armchair and zigzag axes, respectively. Based on previous reports of large anisotropy of calculated and measured electronic and optical properties along these particular directions, in this paper we calculate and compare the EELS spectrum and the intraband low energy plasmon dispersion along $\Gamma-X$ and $\Gamma-Y$, respectively.  

{As reported in the literature, monolayer phosphorene is found to be a direct bandgap semiconductor, the magnitude of the gap being 1.51 eV (0.91 eV), calculated using the HSE06 (GGA-PBE) functional \cite{Qiao:2014, Piyush:2016, PhysRevB.94.205426}.} The GLLB-SC computed electronic bandstructure has a direct bandgap of 1.13 eV and it is shown in Fig~\ref{bs_dos}, along with the corresponding density of states (DOS). 
The conduction band minima (CBM) and the valance band maxima (VBM) are located at the $\Gamma$-point. Interestingly, the electron dispersion is anisotropic and `semi-Dirac' like around this point [see Fig \ref{bs_dos}(b)], having a massive Dirac character along the $\Gamma-X$ direction and parabolic with a large effective mass along the $\Gamma-Y$ direction \cite{2016arXiv161002406D}. This highly anisotropic low energy dispersion of phosphrene is the origin of the direction dependent transport and optical properties of phosphorene. 
 
Pristine phosphorene, being a relatively large bandgap semiconductor, has vanishingly small thermally excited charge carriers even at room temperature and consequently intraband plasmons are absent. In order to explore  intraband plasmons we consider phosphorene doped via electrostatic doping (controlled by a varying gate voltage). For electrostatic doping, the bands near the CBM (and thus the DOS) remains unaffected but the Fermi energy shifts to the conduction band giving rise to doped (electronic) charge carriers, whose number density can be tuned by controlling the gate voltage. 

\begin{figure}
\includegraphics[width=0.90\linewidth]{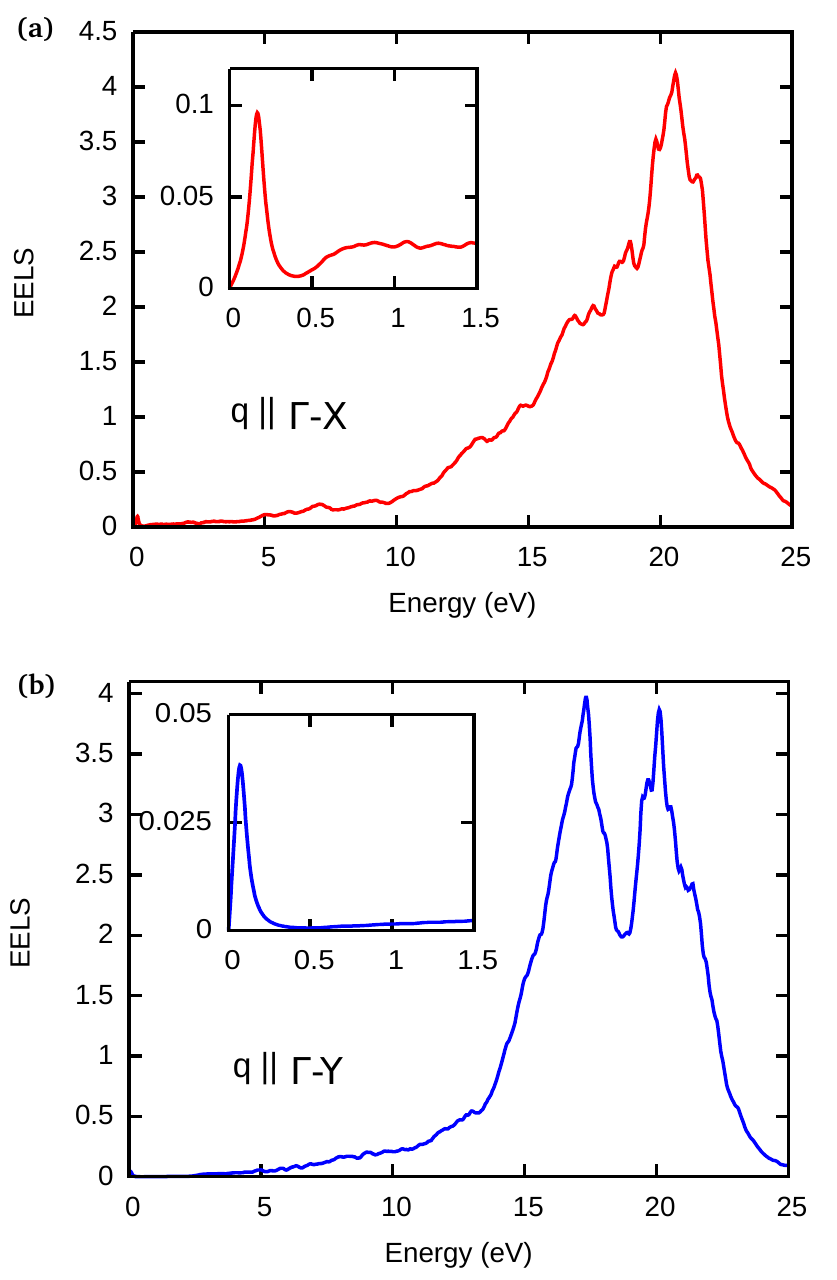}
\caption{EELS (in arbitrary units) of doped bulk black phosphorous along (a) $\Gamma - X$ (armchair) and (b) $\Gamma - Y$ (zigzag) directions as a function of energy for $q = 0.043$ \AA$^{-1}$. The corresponding insets show the low energy behavior. The low energy peaks  for $E \sim 0.35 ~{\rm eV}$ correspond to the intraband plasmon modes of doped black phosphorous. Other high energy peaks correspond to anisotropic excitons ($0.35~{\rm eV} <E < 2~{\rm eV}$) and inter-band plasmonic excitations ($E > 2~{\rm eV}$) -- both of which are identical for the doped as well as the undoped material.  
%{\bf We may later update this figure to a waterfall plot after Barun comes back}.
\label{fig2}}
\end{figure}

The effective low energy Hamiltonian and band-structure of monolayer phosphorene has been derived using  ${\bf k}\cdot {\bf p}$ method\cite{ph-bs3-kp} as well as the tight-binding approach \cite{ph-bs2-tb, ph-bs2-tb-2, ph-bs3-tb, ph-bs4-tb}, both of which yield a qualitatively similar picture \cite{2016arXiv161002406D}. For this manuscript we work with the bare minimum effective low energy Hamiltonian of phosphorene, retaining only the lowest order terms in the wavevectors \cite{1742-6596-603-1-012006},
\be \label{disp0}
H=(uk_y^2+\Delta)\sigma_x+v_{\rm f}k_x\sigma_y~,
\ee
where $\sigma_i$'s are the Pauli matrices. The anisotropic energy spectrum is thus given by
\be \label{disp1}
E_{s}({\bf k})=s\sqrt{v_{\rm f}^2 k_x^2+(uk_y^2+\Delta)^2}~,
\ee
where $s=+1$ $(s=-1)$ corresponds to the conduction (valance) band. Fitting Eq.~\eqref{disp1} to our GLLB-SC dispersion in vicinity of the $\Gamma$ point [see Fig.~\ref{bs_dos}(c)] yields $\Delta= 0.56$ eV, $u = 3.55$ eV/\AA$^2$~ and $v_{\rm f}=4.75$ eV/\AA. Note that, Eq.~\eqref{disp1} can further be approximated as an anisotropic parabolic dispersion given by 
\be \label{disp2}
E_c ({\bf k}) =  E_c+ \frac{\hbar^2 k_x^2}{2 m_x} +  \frac{\hbar^2 k_y^2}{2 m_y}~,
\ee
where $E_c = \Delta$. The values of the anisotropic effective masses for the conduction band are given by $m_x \equiv \hbar^2 \Delta/v_{\rm f}^2 =  0.20 m_e$ and $m_y \equiv \hbar^2/(2u) = 1.1 m_e$, with $m_e$ being the electrons rest mass, consistent with earlier studies \cite{Qiao:2014}. Equations~\eqref{disp0}-\eqref{disp1} will be used to obtain the low energy and low momentum transfer plasmon dispersion of monolayer phosphorene analytically which in turn will be compared with our {\it ab-initio} results.   

\section{Electron energy loss spectrum}
\label{Sec3}
Having discussed the {\it ab-initio} and the low energy electronic band structure, we now proceed to calculate the EELS spectrum of phosphorene. We first test the methodology by calculating the EELS spectrum of bulk black phosphorous and comparing the same with reported data \cite{Schuster,doi:10.1116/1.4926753}. The computed EELS spectrum of bulk black phosphorous is shown in Fig.~\ref{fig2}. As expected, the anisotropic nature of the electronic bandstructure of bulk phosphorous is manifested in the EELS data.  The high energy interband plasmon peak, present in the vicinity of  20 eV in both the $\Gamma-X$ and the $\Gamma-Y$ direction, is in good agreement with the recently published experimental EELS results for bulk black phosphorous \cite{doi:10.1116/1.4926753}. The low energy part of the spectrum, representing intraband plasmons and excitons, are shown in the inset of Fig.~\ref{fig2}, and a marked anisotropy along the two principal direction is observed. In particular, the low energy exciton peak in black phosphorous ($E \approx 0.7$ eV) is only present in the $\Gamma-X$ (armchair) direction, which is consistent with recently reported measurements \cite{Schuster}. Also, the intraband plasmons in the armchair direction ($\Gamma-X$) have a higher intensity as opposed to those in the zigzag ($\Gamma-Y$) direction.   

\begin{figure}
\includegraphics[width=0.90\linewidth]{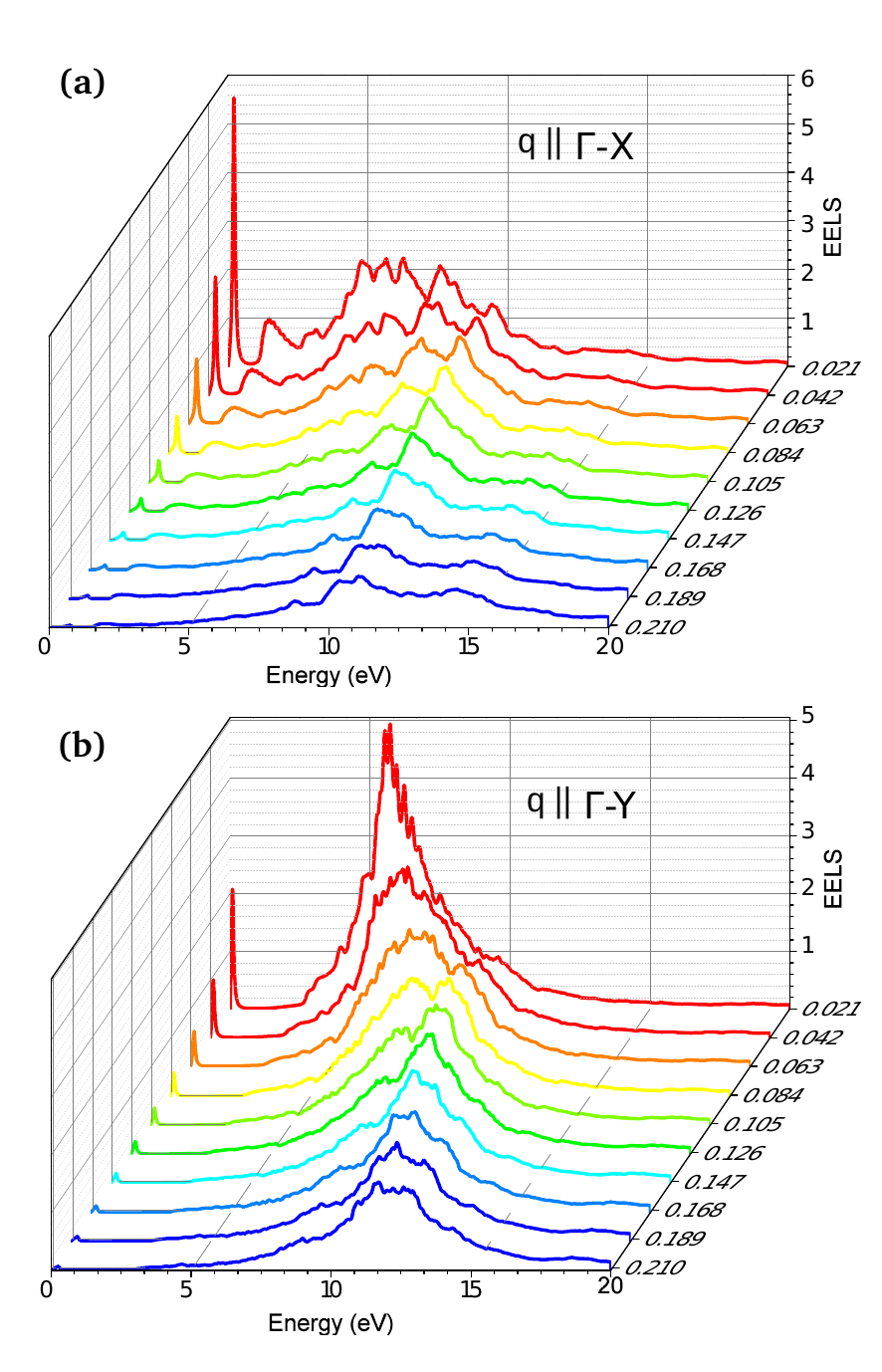}
\caption{EELS (in arbitrary units) of monolayer phosphorene along the (a) $\Gamma - X$ (armchair) and (b) $\Gamma - Y$ (zigzag) directions for different values of momentum transfer, as a function of $q$ and energy (measured from the bottom of the conduction band). The low energy peaks ($E < 1 ~{\rm eV}$) correspond to the intraband plasmons which appear only for the doped case. Other high energy peaks and features corresponding to the excitons and inter-band plasmonic excitations are identical for the doped and the undoped case. %The anisotropy along the $\Gamma - X$ and $\Gamma - Y$ direction in the EELS spectrum for both the intra-band plasmons and other high energy features is evident. 
\label{eels_full}}
\end{figure}

After benchmarking the methodology by successfully reproducing the characteristic features of EELS of bulk black phosphorus, next we focus on monolayer phosphorene. Due to obvious reasons related to it's electronic band structure [see Fig~\ref{bs_dos} (b)], EELS of monolayer phosphorene is also found to be highly anisotropic. This can clearly be observed in Fig.~\ref{eels_full}(a) and (b), where the calculated spectrum of a single layer of doped (by taking $E_{\rm f} = 50$ meV, measured from the CBM) phosphorene is plotted for momentum transfer ($q$) along the $\Gamma -X$ (armchair) and $\Gamma - Y$ (zigzag) directions, respectively. Note that, while the lowest energy peak due to the intraband plasmons are observed only in case of finite doping, others appearing at higher energy have intrinsic origin related to excitons and interband plasmons and they are present in undoped phosphorene as well. Specific features of the EELS are discussed in detail in the following subsections. 

\subsection{Intraband plasmons}
\begin{figure}
\includegraphics[width=0.90\linewidth]{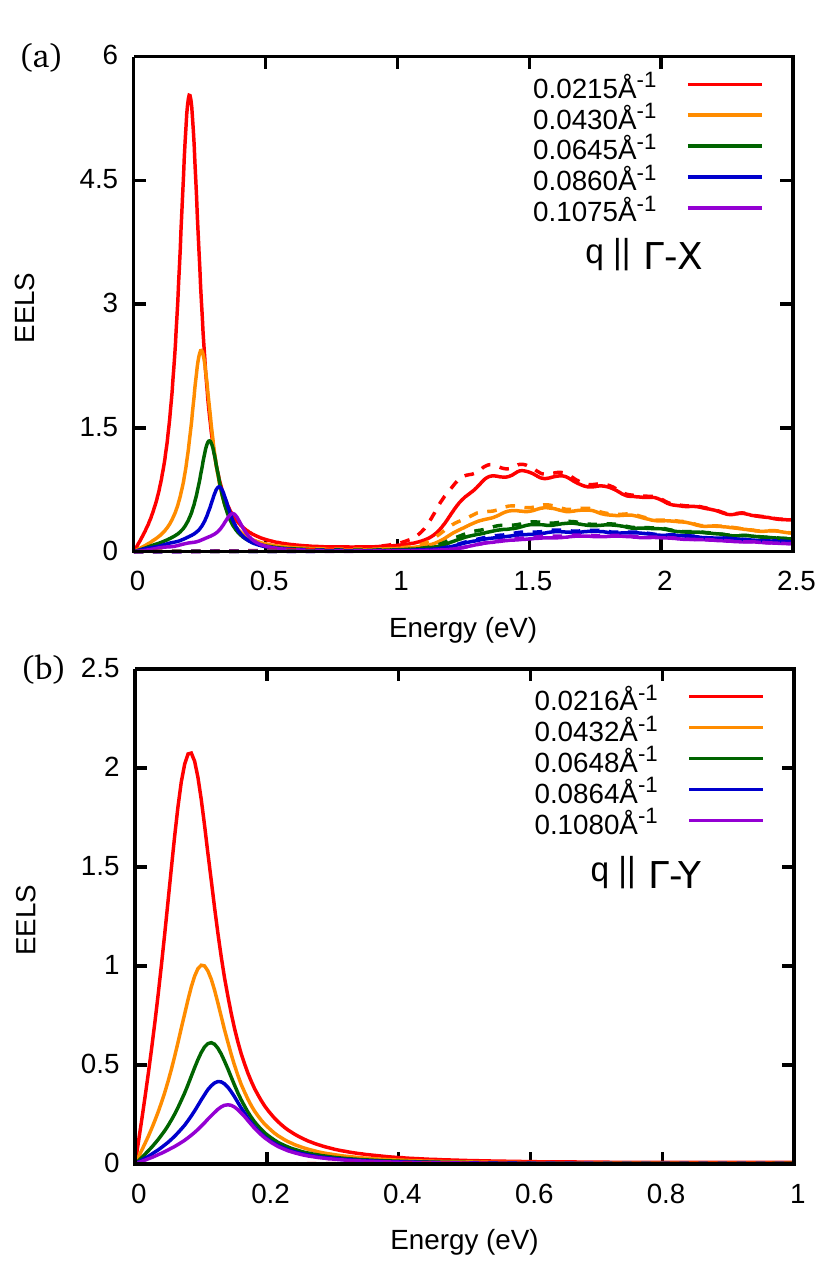}
\caption{Low energy part of the EELS spectrum of monolayer phosphorene for momentum transfer along the (a) $\Gamma - X$ and (b) $\Gamma - Y$ directions. The solid lines denote the doped case with {$E_{\rm f} = 50 $meV}, and the dashed line corresponds to the pristine undoped material. The lowest energy peak ($E <1 $eV) arises from the intraband charge density excitation (plasmon) which are present only in the doped case. The slightly higher energy ($E \sim1$eV) peak which is present only for the momentum transfer along $\Gamma-X$ direction, corresponds to the exciton (electron-hole bound pair). Both the modes are highly dispersive in nature and they gradually loose their strength with increasing momentum transfer. Note that, in panel (b) the dashed lines corresponding to the undoped case, are identically equal to zero due to the absence of the exciton peak.}
	\label{eels_low_energy}
\end{figure}

As shown in Fig.~\ref{eels_full} and Fig.~\ref{eels_low_energy}, the first peak appears for energies less than 0.25 eV. The energy corresponding to the peak is well below the bandgap of the pristine monolayer phosphorene  and in addition this peak is absent in case of undoped phosphorene. Thus we interpret this as the peak originating from the intraband plasmon modes. With increasing momentum transfer, the intensity of intraband plasmons decreases and the peak position shifts to a higher energy, in both $\Gamma-X$ and $\Gamma-Y$ direction [see Fig.~\ref{eels_full} and Fig.~\ref{eels_low_energy}]. Similar type of blue shift is observed as the doping is increased. Alike bulk black phsophorous, the intensity of the intraband peak is lower for momentum transfer along the $\Gamma-Y$ direction, than compared to the $\Gamma-X$ direction, which has the maximum intensity among all the EELS peaks at $q\rightarrow 0$ [see Fig.~\ref{eels_full}]. The anisotropy of intraband plasmon modes and their momentum and doping dependence are discussed in detail in Sec.~\ref{sec:plasmon}.

\subsection{Low energy excitons}
The next peak in the EELS is observed in the vicinity of $1.3$ eV, which is gradually shifted towards the higher energy, accompanied by intensity reduction with increasing $q$. Note that, this peak is much more dispersive compared to the intraband plasmon peaks and extends approximately upto 3 eV (see Fig.~\ref{eels_full} and Fig.~\ref{eels_low_energy}). Since the peak energy coincides approximately with the energy gap of monolayer phosphorene and it exists in undoped phosphorene as well and it is almost independent of doping, we identify this as the lowest energy excitonic peak. As shown in Fig.~\ref{eels_full} and Fig.~\ref{eels_low_energy}, the excitonic peak is highly anisotropic in nature, as it appears only in case of momentum transfer along the $\Gamma-X$ direction, while being completely absent along the $\Gamma-Y$. We believe that the strongly anisotropic optical response reported for monolayer black phsophorene, which is transparent to the incident light in the energy range of 1.1-2.5 eV, but only if it is polarized along the zigzag direction and opaque if the light is polarized in the armchair direction\cite{Tran_2014,Schuster}, originates from the anisotropy of the lowest energy electron-hole pair excitation observed in this work. Note that, since the crystal anisotropy observed in monolayer persists for multilayer, as well as bulk black phosphorous, similar kind of anisotropic excitonic response is expected at higher thickness as well.

\subsection{Interband high energy plasmons}
As we move towards higher energy, the next prominent EELS peak appears in the vicinity of $5$ eV, and similar to the low energy peaks, it's intensity reduces accompanied by a blue shift with increasing momentum transfer. Though it is also anisotropic in nature, this peak has a higher intensity along the zigzag direction as compared to the armchair direction -- unlike the case of intraband plasomons.  As expected, this high energy peak is nearly independent of doping, as it corresponds to very high energy interband transitions.

Comparing the EELS intensity of various peaks as a function of $q$, it is clear that in general the EELS intensity is maximum for direct transitions with $q \to 0$ and decays with increasing $q$. However, the low energy intraband plasmon peak decays more rapidly with increasing $q$ (on account of damping by electron-hole excitations) as compared to the high energy peaks associated with interband transitions. For example, in case of $q\parallel \Gamma-X$, the intraband plasmon peak is the most intense one among all the EELS peaks for $(q=0.021~$\AA$^{-1}$), which almost vanishes at higher $q$, leaving the interband transition peak around 10-14 eV to be the most prominent one. Owing to it's anisotropic nature, the scenario is different in case of $q\parallel \Gamma-Y$, where the interband transition peak around 5 eV has the highest intensity (at $q=0.021~$\AA$^{-1}$) among all the EELS peaks and it broadens significantly and shifts to around 10-14 eV at higher value of $q$.

Note that, other than the intraband plasmons, rest of the peaks corresponding to the interband transitions are likely to be affected due to the bandgap underestimation (approximately 40\%) of GLLB-SC based electronic band structure calculations. For example, while the GW bandgap is reported to be 1.84 eV\cite{PhysRevB.94.205426,Qiao:2014} for monolayer black phosphorus, it is found to be 1.13 eV in our calculation. Thus, the EELS peaks corresponding to interband transitions are expected to be blue shifted in an actual experiment. However, the GLLB-SC bandgap estimation of $0.35$ eV for bulk black phosphorous is very close to its experimentally reported value of $0.31-0.35$ eV \cite{PhysRevB.94.205426}. Thus it turns out that for bulk black phosphorous, EELS peak of 20 eV (as shown in Fig.~\ref{fig2}) based on GLLB-SC calculations is consistent with the 20 eV peak observed in recent experiments\cite{doi:10.1116/1.4926753}. 
\begin{figure}[t]
\includegraphics[width=0.90\linewidth]{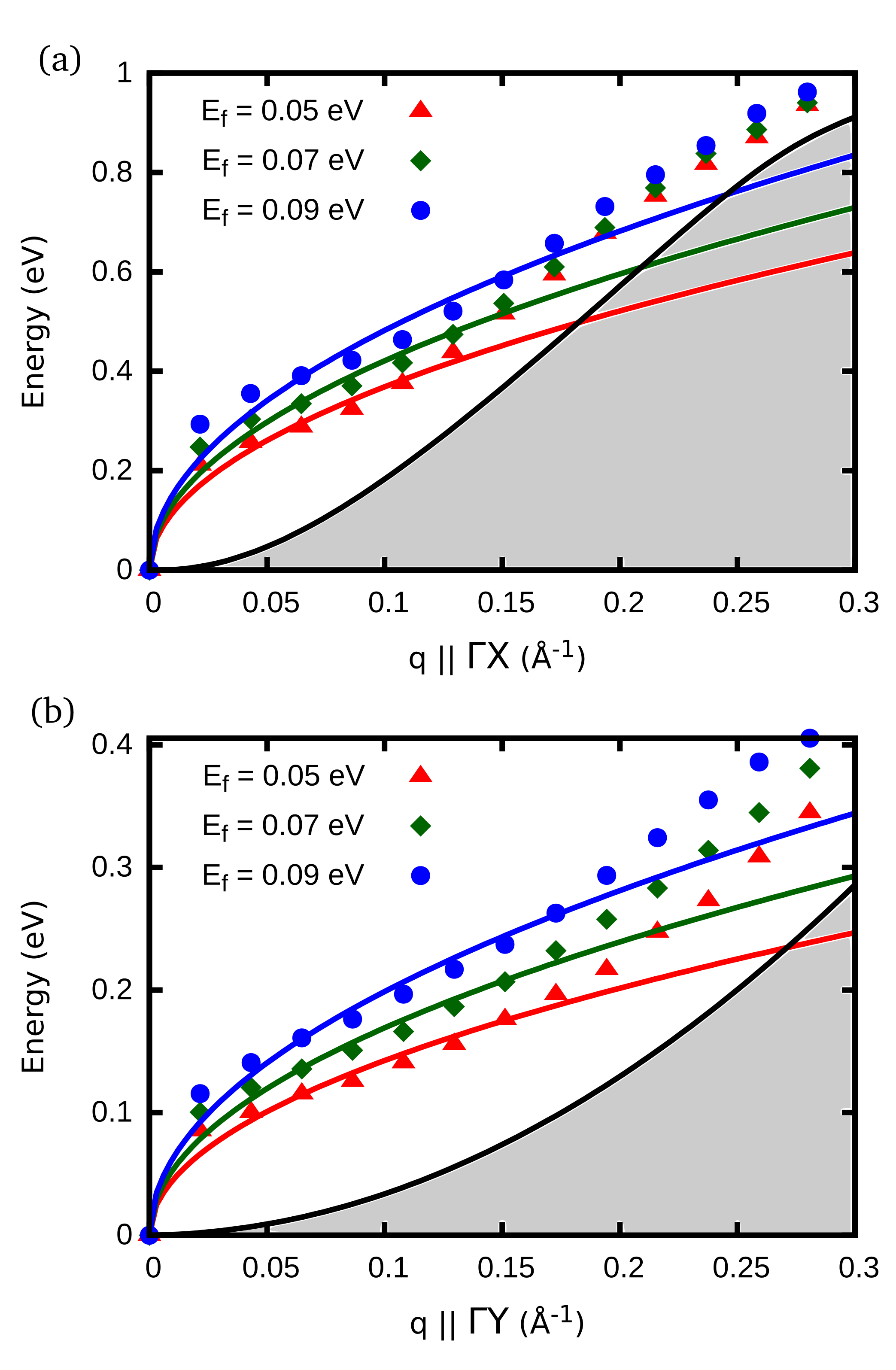}
\caption{Intraband plasmon dispersion of monolayer phosphorene along the (a) $\Gamma - X$ and (b) $\Gamma - Y$ directions at different doping. The corresponding solid lines denote the `universal' long wavelength $\sqrt{q}$ dependence of the plasmon dispersion in two dimensions, which is valid for small $q$ values only. The shaded region in both the panels denotes the single particle continuum (electron-hole excitation spectrum) of the conduction band which is marked by the maxima of  $E_c^{sp} ({\bf q}) = E_c(\Gamma+{\bf q})- E_c(\Gamma)$.} 
\label{Fig5}
\end{figure}
\section{Anisotropic intraband plasmons}
\label{sec:plasmon}
Having described the prominent features of the EELS spectrum of monolayer phosphorene, we now focus on the anisotropic nature of the low energy intraband plasmons and investigate their $q$ and doping ($E_{\rm f}$) dependence. The low energy loss function along the $\Gamma-X$ and $\Gamma-Y$ directions, is shown in Fig.~\ref{eels_low_energy} (a) and (b), respectively. As discussed earlier, the first low energy peak which appears for energies significantly below the bandgap of phosphorene and is only present in case of finite doping, corresponds to intraband plasmons - or collective charge density excitations. As shown in Fig \ref{eels_low_energy}, the intra-band plasmon peak has higher intensity along the $\Gamma-X$ direction as compared to the $\Gamma-Y$ direction, similar to the case of bulk phosphorus. Interestingly, the intensity decay and blue shift of the intraband plasmon peak is more rapid in the  $\Gamma-X$ direction, than compared to the $\Gamma-Y$ direction. 

The momentum dependence of the intraband plasmon peak is further analyzed at different doping (by varying the Fermi energy) in Fig.~\ref{Fig5}. As shown in the figure, for relatively small momentum transfers, the dispersion follows the universal long wavelength $\sqrt{q}$ behavior which is ubiquitous in two dimensional systems \cite{Agarwal2:2015}. For higher value of the momentum transfer $q$, a clear deviation from the $\sqrt{q}$ behavior is observed. Eventually the plasmon dissipates into single particle continuum of the conduction band, whose boundary is marked by the maxima of $E^{sp}_c({\bf q}) = E(\Gamma + {\bf q}) - E(\Gamma)$ [see the shaded area in Fig.~\ref{Fig5}]. As shown in Fig.~\ref{Fig5}, increasing the doping extends the range of validity of $\sqrt{q}$ behavior. For example, at $E_{\rm f}= 50$ meV  the $\sqrt{q}$ fit along the $\Gamma-X$ direction holds good upto $q=0.1$\AA$^{-1}$, which extends upto  $q=0.15$\AA$^{-1}$ with increased doping ($E_{\rm f} = 90$ meV). Similar behaviour of the plasmon dispersion is also seen along the $\Gamma-Y$ direction, albeit the spectral weight of the corresponding plasmon peak is smaller in the $\Gamma-Y$ direction. 

\begin{figure}
\includegraphics[width=0.90\linewidth]{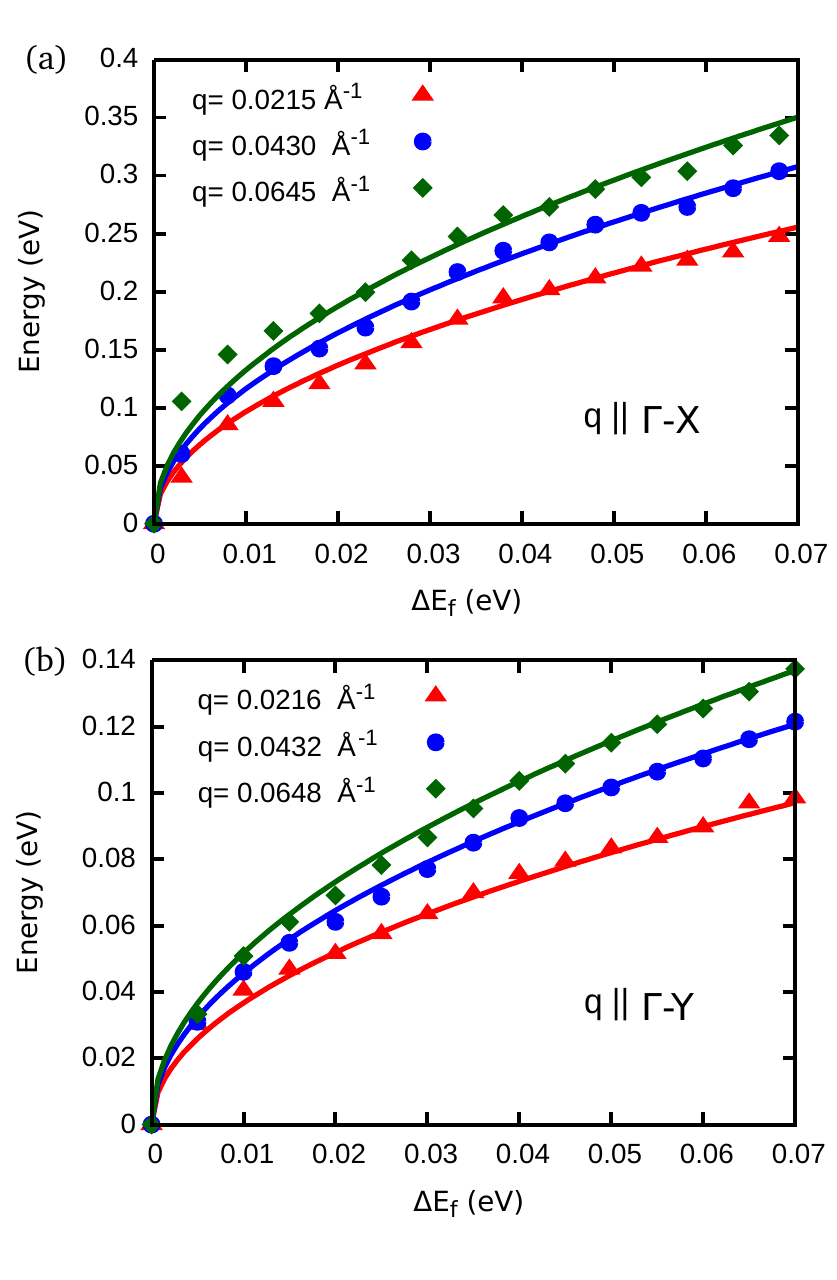}
\caption{Variation of the intraband plasmon energy with the electron doping specified by $\Delta E_{\rm f} = E_{\rm f} - E_c$ along the (a) $\Gamma - X$ and (b) $\Gamma - Y$ directions for different momentum transfer vectors. The doping dependence is of the plasmon dispersion for small $q$ seems to be proportional to $\sqrt{E_{\rm f} -E_c}$, consistent with Eq.~\eqref{pl1}. }
	\label{Fig6}
\end{figure}

\begin{figure}
\includegraphics[width=0.90\linewidth]{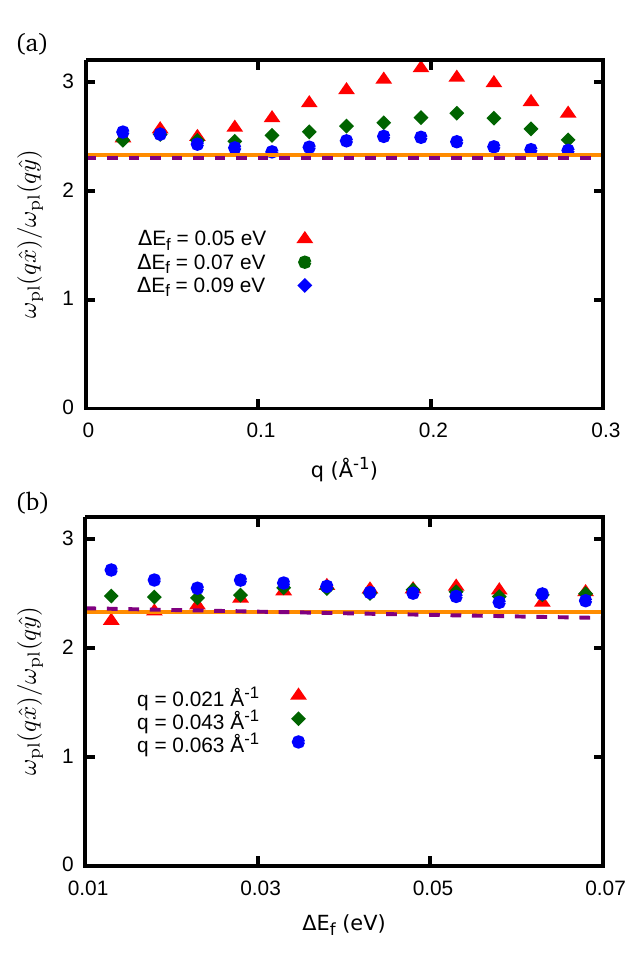}
\caption{Variation of the plasmon anisotropy ratio, $\omega_{\rm pl}(q \hat{x})/\omega_{\rm pl}(q \hat{y})$ with (a) the wavevector $q$ 
for different $E_{\rm f}$ values and (b) the Fermi energy for different $q$. The horizontal straight (orange) line denotes the $\omega_{\rm pl}(q \hat{x})/\omega_{\rm pl}(q \hat{y}) = \sqrt{m_y/m_x} $ line which is completely independent of the doping, based on plasmon dispersion calculated from the anisotropic parabolic dispersion in Eq.~\eqref{pl1}. The dashed horizontal line (violet) denotes Eq.~\eqref{ratio}, which arises from the massive semi Dirac plasmon dispersion, and it depends on the Fermi energy. 
%\textcolor{red}{ In panel a), the dashed line corresponds to which $E_{\rm f}$. Ideally there should be three horizontal lines for three different $E_{\rm f}$. }
}
\label{Fig7}
\end{figure}

Starting from the anisotropic parabolic approximation of the bandstructure of monolayer phosphorene, given by Eq.~\eqref{disp2}, the low energy plasmon dispersion within the random phase approximation,
(for the Coulomb potential $V_q = 2 \pi e^2/\epsilon q$) has been obtained in Ref.~[\onlinecite{Rodin}]. It is explicitly given by 
\be \label{pl1}
\omega_{\rm pl}({\bf q})  = \alpha_0 (E_{\rm f} - E_c)^{1/2} \left[\cos^2 \theta_q  + \frac{m_x}{m_y} \sin^2\theta_q\right]^{1/2}  \sqrt{q}~,
\ee
where we have defined $\theta_q = \tan^{-1}{(q_y/q_x)}$, $\alpha_0^2 = 2 \pi e^2 g_{\rm 2d}/(m_x \epsilon)$, and $g_{\rm 2d} = \sqrt{m_xm_y}/(\pi \hbar^2 )$ is the two dimensional density of states for an anisotropic parabolic band system. 
Eq.~\eqref{pl1} directly yields the following: 1) $\sqrt{q}$ dependence of the plasmon dispersion in all directions for small wavevectors, 2) $\sqrt{E_{\rm f}- E_c}$ dependence of the low energy (and low wave-vector) plasmon dispersion on the Fermi energy, and 
3) the long wave-length anisotropy of the plasmon dispersion $\omega_{\rm pl}(q {\hat x})/\omega_{\rm pl}(q {\hat y}) = \sqrt{m_y/m_x}$, independent of the doping in the system. 

The doping dependence of the intraband plasmon dispersion is shown in Fig.~\ref{Fig6}, and it clearly shows a reasonable match of the plasmon dispersion with the expected $\omega_{\rm pl} \propto \sqrt{E_{\rm f} - E_c}$ dependence of the Fermi energy for small $q$. The plasmon frequency anisotropy ratio is shown in Fig.~\ref{Fig7} and it also seems to be more or less consistent with the $\omega_{\rm pl}(q {\hat x})/\omega_{\rm pl}(q {\hat y}) = \sqrt{m_y/m_x}$ behavior. A more thorough calculation of the long wavelength plasmon dispersion using the semi-Dirac continuum Hamiltonian for phosphorene (see Eq.~\eqref{disp0}), is presented in the appendix, and it also yields qualitatively similar results. However, the anisotropy ratio now explicitly depends on the Fermi energy -- see Eq.~\eqref{ratio}. However we note that this $E_{\rm f}$ dependence of the anisotropic ratio of the plasmon frequency is small -- as shown in panel b) of Fig.~\ref{Fig7}.
 
\section{Summary and conclusions}
\label{Sec5}
In this article we present a thorough study of the anisotropic EELS spectrum of monolayer phosphorene using the TDDFT framework. We find that the anisotropy of the underlaying phosphorene crystal leads to the anisotropy in the band structure and consequently in the EELS spectrum as well -- similar to the case of bulk black phoophorous. For finite doping in the system, the lowest energy peak in the EELS corresponds to the intraband plasmon mode (charge density excitations) in the sub electron volt range.  At slightly higher energy than the band gap, there is a highly dispersive low energy exciton mode, which is almost independent of the doping and it is absent for the momentum transfer along the $\Gamma-Y$ direction. At even higher energies and completely independent of the doping, there are various plasmon modes arising from the interband transitions with a very distinct peak appearing at $\approx 5$eV for monolayer phosphorene (as per GLLB calculations). With increasing number of phosphorene layers (or thickness) this mode is likely to shift to higher energies eventually merging into a 19 eV peak observed in the bulk black phosphorous. We explore the low energy anisotropic intraband plasmons in detail and compare their behavior to analytical expression of the corresponding plasmon dispersion arising from the effective low energy dispersion using RPA. The anisotropic intraband plasmon modes are found to be highly dispersive in nature, with the large wavelength limit following the  $\omega_{\rm pl} \propto \sqrt{q}$ behavior in all directions which is a 
universal characteristic of plasmons in two dimensions.  With increasing doping, the long wavelength plasmon dispersion is found to scale with the Fermi energy as $\omega_{\rm pl} \propto \sqrt{E_{\rm f} - E_c}$. The anisotropy of the long wavelength plasmon dispersion is found to be proportional to the ratio of square root of the effective masses: $\omega_{\rm pl}(q \hat{y})/\omega_{\rm pl}(q \hat{x}) = \sqrt{m_x/m_y}$. 
%We expect these result for the long wavelength intraband plasmons to hold in general for any gapped 2d system. 

\section*{Acknowledgements}
We acknowledge funding from the following DST (Department of science and technology, government of India) schemes: 1) DST INSPIRE Faculty Award, 2)SERB Fast Track Scheme for Young Scientist 3) Ramanujan Fellowship, and 3) DST Nanomission project.
We also thank CC IITK for providing HPC facility.

\appendix
\section{Low energy plasmon dispersion of phosphorene} 
\begin{figure}
\includegraphics[width=0.8\linewidth]{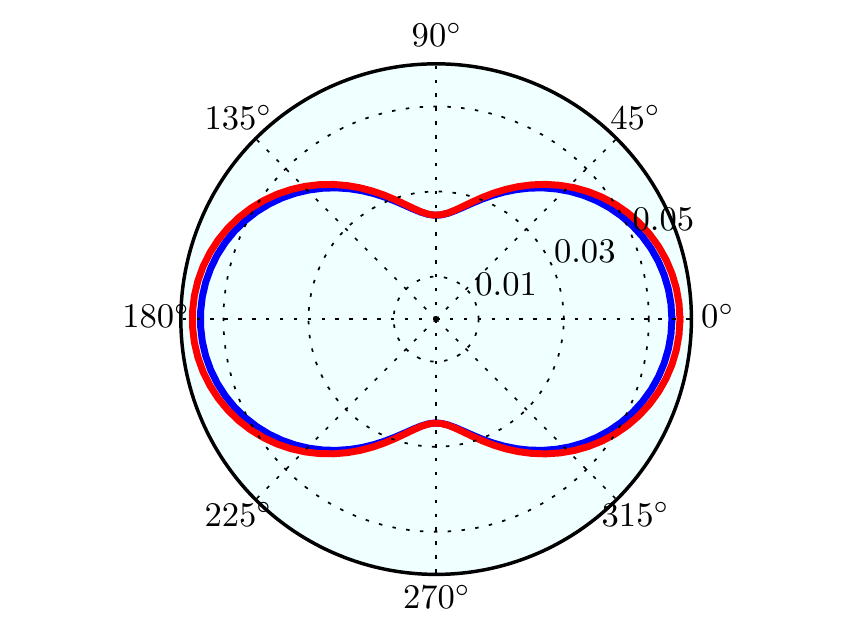}
\caption{Polar plot of the long wavelength plasmon dispersion [in units of $\sqrt{2 e^2/(\hbar^2 \epsilon)}$] of monolayer phosphorene. The blue line is based on Eq.~\eqref{pl1}, and the red line is based on Eq.~\eqref{plasmons_SD}. 
The radial and the azimuthal coordinates are the plasmon frequency, and the direction of the wave vector, where 0 (180) and 90 (270) indicate armchair ($\Gamma-X$) and the zigzag ($\Gamma-Y$) direction, respectively. The momentum is set to be $q= 0.02$~\AA$^{-1}$ and  $E_{\rm f} = 0.07$ eV.}
\label{FigA1}
\end{figure}
Here we calculate the low energy plasmon dispersion for a massive semi Dirac system in two dimensions, described by Eq.~\eqref{disp0}. 
The polarization (density-density response) function is given by 
 \bea \label{polfunc}
\Pi(\vec{q},\omega)& =&\frac{g}{(2\pi)^d}\int d^dk \sum_{ss'=\pm} f^{ss'}({\bf k},{\bf q}) n_{\rm F}(E_{k,s})  \\
&\times &\left(\frac{1}{\hbar\omega^++E_{{\bf k},s}-E_{{\bf k'},s'}}-\frac{1}{\hbar \omega^+ -E_{{\bf k},s}+E_{{\bf k'},s'}}\right), \nn 
\eea
where $\omega^+ = \omega + i \eta$, $E_{k,s}=s\sqrt{v_{\rm f}^2 k_x^2+(uk_y^2+\Delta)^2}$, ${\bf k}'={\bf k}+ {\bf q}$, $f^{ss'}({\bf k},{\bf q})$ is the overlap function of spinors and $n_{\rm F}(E_{k,s})$ is the Fermi function. Expanding the polarization function defined in Eq. (\ref{polfunc}), upto  $q_x^2$ and $q_y^2$ at zero temperature, leads to 
\bea\label{lwexp}
\Pi({\bf q}\to 0,\omega)&=&\frac{g}{4\pi^2}\int_{-\alpha}^{\alpha} {\rm d}k_y \int_{-\beta}^{\beta} {\rm d}k_x   \\
&\times& \Big[ I_1(k_x,k_y)\frac{v_{\rm f}^2 q_x^2}{{\hbar \omega}^2} + I_2(k_x,k_y)\frac{u q_y^2}{{\hbar \omega}^2}\Big], \nn
\eea 
where $\alpha=\sqrt{(E_{\rm f}-\Delta)/u}$, $\beta=v_{\rm f}^{-1}\sqrt{E_{\rm f}^2-(\Delta+uk_y^2)^2}$ and $I_{1,2}(k_x,k_y)$ are complex functions defined as follows
\bea 
I_1&=&\frac{(\Delta+uk_y^2)^2}{[(\Delta+uk_y^2)^2+k_x^2 v_{\rm f}^2]^{3/2}}~,\\
I_2&=&\frac{2[(\Delta+uk_y^2)^3+v_{\rm f}^2k_x^2(\Delta+3uk_y^2)]}{[(\Delta+uk_y^2)^2+k_x^2 v_{\rm f}^2]^{3/2}}~.
\eea 
Performing the integral in Eq. (\ref{lwexp}) we obtain, 
\bea \label{Pi0}
\Pi({\bf q}\to 0,\omega)&=&\frac{g}{4\pi^2 \hbar^2 \omega^2}\Bigg[{v_{\rm f}^2q_x^2}\text{Re}~\zeta_1(E_{\rm f})+{uq_y^2}\nn\\
&\times&\Big[\text{Re}~\zeta_2(E_{\rm f})+\text{Re}~\zeta_3(E_{\rm f})\Big]\Bigg]~,
\eea
where we have defined 
\bea \label{zeta}
\zeta_1(E_{\rm f})&=&-\frac{8i}{3v_{\rm f}E_{\rm f}}\sqrt{\frac{E_{\rm f}-\Delta}{u}}\left[\Delta~G_0(\nu)+E_{\rm f}~G_1(\nu)\right]~,\nn\\
\zeta_2(E_{\rm f})&=&-\frac{32i}{15v_{\rm f}E_{\rm f}}\sqrt{\frac{E_{\rm f}-\Delta}{u}}\Big[(\Delta^2+3E_{\rm f}^2)~G_0(\nu)~,\nn \\
&+&(\Delta-3E_{\rm f})E_{\rm f}~G_1(\nu)\Big]\nn \\
\zeta_3(E_{\rm f})&=&\frac{16iE_{\rm f}}{v_{\rm f}}\sqrt{\frac{E_{\rm f}-\Delta}{u}}\left[G_0(\nu)-G_1(\nu)\right]~. 
\eea  
Equation~\eqref{zeta} in turn uses the following notation,
\bea
\nu&=&\sqrt{\frac{E_{\rm f}-\Delta}{E_{\rm f}+\Delta}},~~~~\phi=i~\text{arcsinh}(\nu) ,\nn\\
G_0(\nu)&=&\text{E}\left[\phi,-\frac{1}{\nu^2}\right],~~~G_1(\nu)=\text{F}\left[\phi,-\frac{1}{\nu^2}\right],
\eea
%polarization function in long wavelength limit reduces to
where $\text{E}\left[\phi,-\frac{1}{\nu^2}\right]$ and $\text{F}\left[\phi,-\frac{1}{\nu^2}\right]$ are incomplete elliptic integral of the first and second kind respectively.

Within RPA, the plasmons modes are given by the zeros of the longitudinal dielectric functions, 
\be \label{plasmon-pole}
\varepsilon(q,\omega)=1-v_q\Pi(q,\omega)=0~,
\ee 
where $v_q$ is the Fourier transform of the Coulomb potential. 
Substituting Eq.~\eqref{Pi0} in Eq.~\eqref{plasmon-pole} immediately yields the long wavelength plasmon dispersion to be 
\bea \label{plasmons_SD}
\omega_{\rm pl} &=& \beta_0 \sqrt{q} \Big[{v_{\rm f}^2 \cos\theta_q^2}~\text{Re}~\zeta_1(E_{\rm f})+{u 
\sin\theta_q^2}\nn\\
&\times&\left(\text{Re}~\zeta_2(E_{\rm f})+\text{Re}~\zeta_3(E_{\rm f})\right)\Big]^{1/2}~,
\eea
where we have used the two dimensional form of $V_q = 2 \pi e^2/(\epsilon q)$ and defined $\beta_0^2 = g e^2/(2\pi \hbar^2 \epsilon)$. 
Figure~\ref{FigA1}, shows the angular dependence of the long wavelength plasmon frequency, specified by Eq.~\eqref{plasmons_SD}, and it matches reasonably well with the long wavelength expression given in Eq.~\eqref{pl1}.

The ratio of the anisotropic plasmon dispersion in $x$ and $y$ direction for same value of the wave vector (in the long wavelength regime) is given by
\be \label{ratio}
\frac{\omega_{\rm pl}(q \hat{x})}{\omega_{\rm pl}(q \hat{y})} =\sqrt{ \frac{v_{\rm f}^2~\text{Re}~\zeta_1(E_{\rm f})}{u[\text{Re}~\zeta_2(E_{\rm f})+\text{Re}~\zeta_3(E_{\rm f})]}}~.
\ee

\bibliography{ELSS_phos}

\end{document}